\documentstyle[12pt]{article}
\input epsf.tex
\catcode`\@=11

\@addtoreset{equation}{section}
\def\theequation{\arabic{section}.\arabic{equation}}

\catcode`\@=11

\def\section{\@startsection{section}{1}{\z@}{3.5ex plus 1ex minus
   .2ex}{2.3ex plus .2ex}{\large\bf}}

\newskip\humongous \humongous=0pt plus 1000pt minus 1000pt

\newif\ifdtup

\def\eqnarray{\let\@currentlabel=\theequation\refstepcounter{equation}
    \global\@eqnswtrue
    \global\@eqcnt\z@\tabskip\@centering\let\\=\@eqncr
    $$\halign to \displaywidth\bgroup\@eqnsel\hskip\@centering
      $\displaystyle\tabskip\z@{##}$&\global\@eqcnt\@ne
       \hfil${{}##{}}$\hfil
      &\global\@eqcnt\tw@ $\displaystyle\tabskip\z@{##}$\hfil
       \tabskip\@centering&\llap{##}\tabskip\z@\cr}
\def\lefteqn#1{\hbox to 4\arraycolsep{$\displaystyle #1$\hss}}
\def\thesection{\arabic{section}.}

\def\appendix{\setcounter{section}{0}
        \def\thesection{Appendix.}
        \def\theequation{\Alph{section}.\arabic{equation}}}

\long\def\@makefntext#1{\parindent 0cm\noindent
\hbox to 1em{\hss$^{\@thefnmark}$}#1}
\def\IR{{\hbox{{\rm I}\kern-.2em\hbox{\rm R}}}}
\def\IH{{\hbox{{\rm I}\kern-.2em\hbox{\rm H}}}}
\def\IC{{\ \hbox{{\rm I}\kern-.6em\hbox{\bf C}}}}
\def\IZ{{\hbox{{\rm Z}\kern-.4em\hbox{\rm Z}}}}
\def\rref#1{(\ref{#1})}
\def\zx{{\vphantom{-1}}}
\newcommand{\beq}{\begin{equation}}
\newcommand{\eeq}{\end{equation}}
\newcommand{\NPB}[1]{{\sl Nucl.~Phys.}~{\bf B#1}}
\newcommand{\Ann}[1]{{\sl Ann.~Phys.}~{\bf #1}}
\newcommand{\CMP}[1]{{\sl Commun.~Math.~Phys.}~{\bf #1}}

\newcommand{\CQG}[1]{{\sl Class.~Quant.~Grav.}~{\bf #1}}

\newcommand{\JMP}[1]{{\sl J.~Math.~Phys.}~{\bf #1}}
\begin{document}
%
%
%
%
\def\citen#1{%
\edef\@tempa{\@ignspaftercomma,#1, \@end, }
\edef\@tempa{\expandafter\@ignendcommas\@tempa\@end}%
\if@filesw \immediate \write \@auxout {\string \citation {\@tempa}}\fi
\@tempcntb\m@ne \let\@h@ld\relax \let\@citea\@empty
\@for \@citeb:=\@tempa\do {\@cmpresscites}%
\@h@ld}
%
\def\@ignspaftercomma#1, {\ifx\@end#1\@empty\else
   #1,\expandafter\@ignspaftercomma\fi}
\def\@ignendcommas,#1,\@end{#1}
%
%
\def\@cmpresscites{%
 \expandafter\let \expandafter\@B@citeB \csname b@\@citeb \endcsname
 \ifx\@B@citeB\relax 
    \@h@ld\@citea\@tempcntb\m@ne{\bf ?}%
    \@warning {Citation `\@citeb ' on page \thepage \space undefined}%
 \else
    \@tempcnta\@tempcntb \advance\@tempcnta\@ne
    \setbox\z@\hbox\bgroup 
    \ifnum\z@<0\@B@citeB \relax
       \egroup \@tempcntb\@B@citeB \relax
       \else \egroup \@tempcntb\m@ne \fi
    \ifnum\@tempcnta=\@tempcntb 
       \ifx\@h@ld\relax 
          \edef \@h@ld{\@citea\@B@citeB}%
       \else 
          \edef\@h@ld{\hbox{--}\penalty\@highpenalty \@B@citeB}%
       \fi
    \else   
       \@h@ld \@citea \@B@citeB \let\@h@ld\relax
 \fi\fi%
 \let\@citea\@citepunct
}
%
\def\@citepunct{,\penalty\@highpenalty\hskip.13em plus.1em minus.1em}%
%
%
\def\@citex[#1]#2{\@cite{\citen{#2}}{#1}}%
%
%
\def\@cite#1#2{\leavevmode\unskip
  \ifnum\lastpenalty=\z@ \penalty\@highpenalty \fi 
  \ [{\multiply\@highpenalty 3 #1
      \if@tempswa,\penalty\@highpenalty\ #2\fi 
    }]\spacefactor\@m}
\let\nocitecount\relax  
%
\begin{titlepage}
\vspace{.5in}
\begin{flushright}
UCD-11-94\\
gr-qc/9406006\\
April 1994\\
\end{flushright}
\vspace{.5in}
\begin{center}
{\Large\bf
 Topology Change\\[1ex] in (2+1)-Dimensional Gravity}\\
\vspace{.4in}
{S.~C{\sc arlip}\footnote{\it email: carlip@dirac.ucdavis.edu}\\
       {\small and}\\
R.~C{\sc osgrove}\footnote{\it email: cosgrove@bethe.ucdavis.edu}\\
       {\small\it Department of Physics}\\
       {\small\it University of California}\\
       {\small\it Davis, CA 95616}\\{\small\it USA}}
\end{center}

\vspace{.5in}
\begin{center}
\begin{minipage}{5in}
\begin{center}
{\large\bf Abstract}
\end{center}
{\small
In (2+1)-dimensional general relativity, the path integral for a
manifold $M$ can be expressed in terms of a topological invariant,
the Ray-Singer torsion of a flat bundle over $M$.  For some manifolds,
this makes an explicit computation of transition amplitudes possible.
In this paper, we evaluate the amplitude for a simple topology-changing
process.  We show that certain amplitudes for spatial topology change
are nonvanishing---in fact, they can be infrared divergent---but
that they are infinitely suppressed relative to similar
topology-preserving amplitudes.
}
\end{minipage}
\end{center}
\end{titlepage}
\addtocounter{footnote}{-2}

\section{Introduction}

The path integral in general relativity is a sum over geometries, and
it is natural to ask whether this sum should be extended to include
topologies as well \cite{Wheeler}.  This question can take two forms:
(1) should path integrals include sums over intermediate {\em spacetime}
topologies (``spacetime foam''); and (2) should we allow transitions
between different {\em spatial} topologies?  These issues are closely
related---it is hard to imagine a formalism that permits spatial
topology change while forbidding sums over intermediate topologies---but
they are distinct.  In this paper, we focus on the second.

Topology change, if it occurs, is expected to be a quantum mechanical
process \cite{Geroch}.   Unfortunately, we do not yet have a viable
quantum theory of gravity with which to compute topology-changing
amplitudes.  We can, however, look for models that give us hints
about what to expect in the full theory.

One particularly interesting model is (2+1)-dimensional gravity,
standard general relativity in two spatial dimensions plus time.
The underlying conceptual issues of quantum gravity, and some of the
technical aspects as well, are identical in 2+1 and 3+1 dimensions.
But the elimination of one dimension vastly simplifies the theory,
making many computations possible for the first time.  Moreover,
general relativity in 2+1 dimensions is renormalizable (it is
in fact finite), allowing us to avoid the difficult problems of
interpreting path integrals in (3+1)-dimensional gravity.

Indeed, as first noted by Witten \cite{Wit1,Wit2}, the path integral
for (2+1)-dimensional gravity can be evaluated explicitly in terms
of a standard topological invariant, the Ray-Singer torsion.  In
first-order form, the action may be written schematically as
\beq
S = \int_M   e^a R_a, \qquad
  R_a = d\omega_a + {1\over2}\epsilon_{abc}\omega^b\omega^c
\label{a1}
\eeq
(see the next section for details).  The Lagrangian is cubic in the
fields, but transition amplitudes can still be computed nonperturbatively:
the triad $e$ acts as a Lagrange multiplier, giving a delta functional
that absorbs the remaining integral over $\omega$.  With careful
gauge-fixing, one obtains a combination of determinants of Laplacians
known as the Ray-Singer torsion, a well-known topological invariant of
a flat bundle over $M$.  This torsion, in turn, is equivalent to the
Reidemeister torsion, a combinatorial invariant that can be relatively
simple to compute.

In reference \cite{Wit2}, Witten points out that this equality makes
the explicit computation of topology-changing amplitudes possible.  He
obtains a set of selection rules relating holonomies of $\omega$ on the
initial and final boundaries of $M$.  In reference \cite{Amano}, Amano
and Higuchi obtain a much more stringent set of selection rules by
demanding that the initial and final boundaries be spacelike.  They
further write down a set of three-manifolds, expressed in terms of
Heegaard decompositions, that interpolate between all topology changes
allowed by their selection rules among surfaces of genus greater than
one.

The initial goal of this paper was to compute the Reidemeister torsion
for the simplest of these interpolating three-manifolds, thus obtaining
an explicit topology-changing amplitude.  We do so in section 6.  In
the course of the computation, however, we found it necessary to
reanalyze Witten's expression for the path integral, treating boundary
conditions somewhat more carefully.  To our surprise, we discovered that
the path integrals for the Amano-Higuchi manifolds typically involve
zero-modes of the triad $e$, leading to infrared divergences of the
type discussed by Witten in ``classical'' spacetimes \cite{Wit2}.

To make sense of our topology-changing amplitudes, we must therefore add
an infrared cutoff and compare the results to similar topology-preserving
amplitudes.  Unfortunately, the meaning of the word ``similar'' is not
entirely clear.  As noted above, once we permit spatial topology change,
we ought to also allow sums over intermediate spacetime topologies as well.
Such sums are typically badly divergent \cite{CarDe}, and it is not clear
how to normalize the total amplitude.  This is a deep conceptual question,
which goes beyond the issue of divergences: a topology-changing amplitude
relates states in different Hilbert spaces, and we do not yet have a
``Fock space'' that includes multiple topologies.

We shall not attempt to address this issue here.  Instead, we take
the less ambitious route of normalizing our amplitude relative
to a topology-preserving amplitude coming from a three-manifold with
a related Heegaard decomposition.  We shall show that the effect of the
infrared divergences discussed above is to infinitely suppress the
topology-changing amplitude.  In the absence of a clear prescription
for normalization, we cannot yet claim that quantum gravity prohibits
topology change, but we find this result rather suggestive.

\section{Action, States, and Boundary Conditions}

We begin with a brief review of (2+1)-dimensional gravity in the
first-order formalism \cite{Wit1}.  Let $M$ be a three-dimensional
spacetime.  As our fundamental variables, we choose a triad
$e_\mu^{\ a}$---a section of the bundle of orthonormal frames on $M$---and
an SO(2,1) connection on the same bundle, which we describe by a connection
one-form $\omega_\mu{}^a{}_b$.  The standard Einstein-Hilbert action is
\beq
I_{\hbox{\scriptsize grav}}[M] = \int_M\,e^a\wedge
  \left(d\omega_a+{1\over2}\epsilon_{abc}\,\omega^b\wedge\omega^c\right),
\label{b1}
\eeq
where $e^a = e_\mu{}^a dx^\mu$ and $\omega^a = {1\over2}\epsilon^{abc}
\omega_{\mu bc}dx^\mu$.  The equations of motion coming from this action
are easily derived: they are
\beq
de^a + \epsilon^{abc}\omega_b\wedge e_c = 0
\label{b2}
\eeq
and
\beq
R^a = d\omega^a + {1\over2}\epsilon^{abc}\omega_b\wedge\omega_c = 0 .
\label{b3}
\eeq
The first of these is the condition of vanishing torsion, which ensures
that the connection $\omega$ is compatible with the metric $g_{\mu\nu} =
\eta_{ab}e_\mu{}^ae_\nu{}^b$.  The second is then equivalent to the
ordinary vacuum field equations of general relativity.  Note that \rref{b3}
can also be interpreted as the requirement that the connection $\omega$ be
flat, a feature peculiar to 2+1 dimensions that goes a long way towards
explaining the model's simplicity.

The action \rref{b1} is invariant under local SO(2,1) transformations,
\begin{eqnarray}
\delta e^a &=& \epsilon^{abc}e_b\tau_c \nonumber\\
\delta \omega^a &=& d\tau^a + \epsilon^{abc}\omega_b\tau_c ,
\label{b4}
\end{eqnarray}
as well as ``local translations,''
\begin{eqnarray}
\delta e^a &=& d\rho^a + \epsilon^{abc}\omega_b\rho_c \nonumber\\
\delta \omega^a &=& 0 .
\label{b5}
\end{eqnarray}
These transformations together form an ISO(2,1) algebra, and Witten
has shown that the one-forms $e^a$ and $\omega^a$  constitute an
ISO(2,1) connection.  $I_{\hbox{\scriptsize grav}}$ is also invariant
under diffeomorphisms, but this is not an independent symmetry: when
the triad $e_\mu{}^a$ is invertible, diffeomorphisms in the connected
component of the identity are equivalent to transformations of the
form \rref{b4}--\rref{b5}.  We must still account for the ``large''
diffeomorphisms---the mapping class group of $M$---but for most of
this paper, these will not play an important role.

Since we are interested in topology change, we shall use path integral
techniques to quantize the action \rref{b1}.  But because we are dealing
with manifolds with boundary, we must first determine the appropriate
boundary conditions.  A simple heuristic argument is as follows.  Let
us start by picking boundary values for either $e$ or $\omega$ on
$\partial M$.  This amounts to choosing Dirichlet boundary conditions
for (say) $\omega$.  Now, the kinetic term in the action \rref{b1} can
be written as $<\!e,*d\omega\!>$, where $*$ is the Hodge star operator
and $<,>$ is the corresponding inner product.  But if $\omega$ obeys
Dirichlet boundary conditions, $*d\omega$ obeys Neumann boundary
conditions, so we ought to require the same of $e$ in the inner
product.  We thus expect opposite boundary conditions for our two
fields.

To make this argument more rigorous, let us first examine the Hilbert
space of (2+1)-dimensional quantum gravity.  Canonical quantization of
this theory on a manifold $\IR\!\times\!\Sigma$ has been discussed by a
number of authors; see \cite{CarSix} for a summary.  A key feature is
that the classical reduced phase space is a cotangent bundle, whose
base space is the space of flat connections $\omega$ on a slice $\Sigma$
modulo SO(2,1) gauge transformations and large diffeomorphisms.  In the
simplest approach to quantization, the ``connection representation''
\cite{Ash}, states are therefore gauge-invariant functionals
$\Psi[\omega_i{}^a]$ of the spatial part of the connection, subject
to the constraint that $\omega$ be flat on $\Sigma$.

The corresponding boundary conditions for the path integral therefore
require us to to fix a flat connection $\omega_i{}^a$ on $\partial M$.
More precisely, let $I\!:\partial M \!\hookrightarrow\! M$ be the
inclusion map.  We can then freely specify the induced connection
one-form $I^*\omega$ on $\partial M$, as long as the induced curvature
$I^*R$ vanishes.  SO(2,1) gauge invariance of the resulting amplitude
is formally guaranteed by the functional integral over the normal
component of $\omega$: at $\partial M$,  $\omega_{\perp a}$ is a
Lagrange multiplier for the constraint
\beq
N^a = {1\over2}\epsilon^{ij}\left(
 \partial_i e_j{}^a - \partial_j e_i{}^a
 + \epsilon^{abc}(\omega_{ib}e_{jc} - \omega_{ic}e_{jb})\right)
\label{b5b}
\eeq
that generates SO(2,1) transformations of $I^*\omega$ \cite{Wit1,Carscat}.
Observe that we must integrate over $\omega_\perp$ at the boundary to
enforce this constraint---that is, we must not fix $\omega_\perp$ as
part of the boundary data.  This is in accord with the canonical theory,
in which wave functionals depend only on the tangential components of
$\omega$.

Note that the flat connection $I^*\omega$ is completely determined by
its holonomies, that is, by a group homomorphism
\beq
H: \pi_1(\partial M) \rightarrow \hbox{SO(2,1)} .
\label{b6}
\eeq
As we shall see later, the transition amplitude can be described rather
explicitly in terms of these holonomies.

The specification of $I^*\omega$ is not quite sufficient to give us a
well-defined path integral.  To obtain an additional boundary condition,
it is useful to decompose $\omega$ into a background field $\bar\omega$
that satisfies the classical field equations and a fluctuation $\Omega$:
\beq
\omega = \bar\omega + \Omega , \qquad
 d\bar\omega^a+{1\over2}\epsilon^{abc}\bar\omega_b\wedge\bar\omega_c = 0.
\label{b7}
\eeq
Note that if there is no classical field with our chosen boundary values,
then the transition amplitude is zero: the integral over $e$ gives a
delta functional $\delta[R^a]$ that vanishes everywhere.  The requirement
of existence of a classical solution leads to Witten's selection rules for
topology change \cite{Wit2}.  Assuming now that $\bar\omega$ exists,
the boundary condition $I^*\Omega=0$ can be recognized as a part of
the standard Dirichlet, or relative, boundary conditions for a one-form
\cite{RaySinger,Schw},
\beq
I^*\Omega = 0 = *{\bar D}\!*\!\Omega .
\label{b8}
\eeq
Here, $*$ is the Hodge star operator with respect to an auxiliary
Riemannian metric $h$ that we introduce in order to define a direction
normal to the boundary, while ${\bar D}$ is the covariant exterior
derivative coupled to the background connection $\bar\omega$,
$$ {\bar D}\beta^a
 = d\beta^a + \epsilon^{abc}{\bar\omega}_b\wedge\beta_c.$$
These boundary conditions make the operators $\bar D$ and $*{\bar D}*$
hermitian conjugates, and guarantee that the Laplacian ${\bar\Delta} =
{\bar D}*{\bar D}* + *{\bar D}*{\bar D}$ is hermitian.  We shall see below
that the new condition on the derivatives of $\Omega$, which can be written
in component form as $\bar D_\mu\Omega^\mu|_{\partial M}=0$, is actually
a gauge condition.  Note that \rref{b8} depends on the nonphysical metric
$h$; we must check that the final transition amplitudes are independent
of $h$.

We next turn to the boundary conditions for the triad $e$.  $I^*e$
and $I^*\omega$ are conjugate variables, so we cannot expect to specify
them simultaneously.  On the other hand, we must {\em not} integrate over
the normal component of $e$ at the boundary.  Indeed, $e_{\perp a}$ acts
as a Lagrange multiplier for the constraint
\beq
\tilde N^a = {1\over2}\epsilon^{ij}\left(
 \partial_i \omega_j{}^a - \partial_j \omega_i{}^a
 + \epsilon^{abc}\omega_{ib}\omega_{jc} \right) ,
\label{b5c}
\eeq
and would lead to a delta functional $\delta[\tilde N^a]=\delta[I^*R^a]$
at the boundary.  But we have already required that $I^*\omega$ be flat,
so such a delta functional would diverge.  We  avoid this redundancy by
fixing $e_\perp$ at $\partial M$.  We shall see below that transition
amplitudes do not depend on the specific value of $e_\perp$, so this does
not contradict the canonical picture.

As with $\omega$, we can obtain additional boundary conditions by
decomposing $e$ into a classical background field and a fluctuation
\beq
e = \bar e + E , \qquad d\bar e^a
 + \epsilon^{abc}\bar\omega_b\wedge \bar e_c = 0
\label{b9}
\eeq
where $E_\perp$ vanishes, i.e., $I^*(*E)=0$.  This restriction on $E$ is
a part of the standard Neumann, or absolute, boundary conditions for a
one-form,
\beq
I^*(*E) = 0 = I^*(*{\bar D}E) .
\label{b10}
\eeq
Once again, these conditions make the Laplacian $\bar\Delta$ hermitian.

\section{Path Integrals and Ray-Singer Torsion\protect\footnotemark}
\footnotetext{Much of this section is a
summary and elaboration of Witten's work in \cite{Wit2}.}

We are interested in path integrals of the form
\beq
Z_M[I^*\omega] = \int [d\omega][de]\,
 \exp\left\{ iI_{\hbox{\scriptsize grav}}[M] \right\} ,
\label{c1}
\eeq
where $M$ is a manifold whose boundary
\beq
\partial M = \Sigma_1 \amalg \Sigma_2
\label{c2}
\eeq
is the disjoint union of an ``initial'' surface $\Sigma_1$ and a ``final''
surface $\Sigma_2$.  ($\Sigma_1$ and $\Sigma_2$ need not be connected
surfaces.)  Our first step is to choose gauge conditions to fix the
transformations \rref{b4}--\rref{b5}.  To do so, we employ the auxiliary
Riemannian metric $h$ introduced in the last section, and impose the
Lorentz gauge conditions
\beq
*\!D\!*\!E^a = *D\!*\!\Omega^a = 0 .
\label{c3}
\eeq
For later convenience, we use the covariant derivative $D$ coupled to
the full connection $\omega$ rather than $\bar\omega$ in our gauge-fixing
condition.  $D$ and $\bar D$ agree at the boundary, however, so the gauge
condition on $\Omega$ reduces to the second equation of \rref{b8} on
$\partial M$, as promised.

To impose \rref{c3} in the path integral, we introduce a pair of
three-form Lagrange multipliers $u_a$ and $v_a$, and add a term
\beq
I_{\hbox{\scriptsize gauge}} = -\int_M \left( u_a\wedge *D\!*\!E^a
  + v_a\wedge* D\!*\!\Omega^a \right)
\label{c4}
\eeq
to the action.  It is not hard to see that for the path integral to be
well-defined, $u$ should obey relative boundary conditions ($I^*(*D\!*\!u)
= 0$), while $v$ should obey absolute boundary conditions ($*v=0$ on
$\partial M$).  The latter restriction again has a rather straightforward
interpretation: since we are already imposing the gauge condition \rref{b8}
on $\Omega$ at the boundary, we do not need the added delta functional
$\delta[*D\!*\!\Omega]$ that would come from integrating over $v$ at
$\partial M$.

As usual, the process of gauge-fixing leads to a Faddeev-Popov
determinant, which can be incorporated by adding a ghost term
\beq
I_{\hbox{\scriptsize ghost}} = -\int_M \left( {\bar f}\wedge *D\!*\!D f +
  {\bar g}\wedge *D\!*\!D g \right) ,
\label{c5}
\eeq
where $f$, $\bar f$, $g$, and $\bar g$ are anticommuting ghost fields.
We must again be careful about boundary conditions: corresponding to
restrictions \rref{b8} and \rref{b10} on $\Omega$ and $E$, we choose
$f$ and $\bar f$ to satisfy relative boundary conditions and $g$ and
$\bar g$ to satisfy absolute boundary conditions.  The full gauge-fixed
action is then
\begin{eqnarray}
I &=& I_{\hbox{\scriptsize grav}} + I_{\hbox{\scriptsize gauge}} +
  I_{\hbox{\scriptsize ghost}} \nonumber\\
&=& \int_M \Bigl[
  E^a\wedge\left({\bar D}\Omega_a
  + {1\over2}\epsilon_{abc}\,\Omega^b\wedge\Omega^c
  + *D\!*\!u_a\right) \nonumber\\
  &\phantom{=}& +\
  {1\over2}\epsilon_{abc}\,{\bar e}^a\wedge\Omega^b\wedge\Omega^c
  - v^a\wedge*D*\Omega_a
  - {\bar f}\wedge *D\!*\!D f - {\bar g}\wedge *D\!*\!D g \Bigr] .
\label{c6}
\end{eqnarray}

$E$ and $v$ occur linearly in \rref{c6}, so following Witten
\cite{Wit2}, we shall first integrate over these fields to obtain delta
functionals.  There is one subtlety here: certain modes of $E$ do not
contribute to the action, and must be treated separately in the
integration measure.  The issue is most easily understood in the case of
a linear action, for instance
$$I = \int_M d^3x\sqrt{g}\, \alpha\Delta\beta .$$
If we expand $\alpha$ and $\beta$ in terms of orthonormal modes $\phi_n$
of the Laplacian, $\alpha = \sum a_n\phi_n$, $\beta = \sum b_n\phi_n$,
it is easy to see that $I = \sum{}' \lambda_na_nb_n$, where $\lambda_n$
are the eigenvalues of the Laplacian and the sum automatically excludes
the coefficients of the zero-modes of $\Delta$.  The path integral
measure $[d\alpha]$ thus splits into an integral over modes for which
$\lambda_n\ne0$---giving a delta function $\prod{}'\lambda_n^{-1}
\delta(b_n)$---and an integral $\int[da_0]$ over zero-modes.

Unfortunately, the action \rref{c6} is not linear in $\Omega$, and such a
mode decomposition is much more difficult.\footnote{See \cite{Blau} for a
related treatment of the Abelian $B$-$F$ system, in which this nonlinearity
is not an issue.} However, we can argue as follows.  The integral over
the ``nonzero-modes'' of $E$ will give a delta functional of ${\bar D}
\Omega_a + {1\over2}\epsilon_{abc}\Omega^b\wedge\Omega^c + *D\!*\!u_a$.
The zeros of this expression form a surface $(\tilde\Omega(s), \tilde u(s))$
in the space of fields, and if we expand the action around these zeros,
only those fields infinitesimally close to this surface should contribute
to the path integral.  Writing $\Omega = \tilde\Omega + \delta\Omega$, we
easily find that the relevant zero-modes of $E$ are those $\tilde E$
for which
\beq
D_{\bar\omega + \tilde\Omega}\tilde E = 0 .
\label{c6a}
\eeq
(Note that the $\tilde E$ depend on $\tilde\Omega$, so the order of
integration below cannot be changed.)  Performing the integration
over $E$ and $v$, we obtain
\beq
\int [d\Omega][du][dE][dv] e^{iI} =
\int [d\Omega][du][d\tilde E]
  \delta\bigl[{\bar D}\Omega_a
  + {1\over2}\epsilon_{abc}\,\Omega^b\wedge\Omega^c
  + *D\!*\!u_a\bigr] \delta\bigl[*D\!*\!\Omega_a\bigr] .
\label{c7}
\eeq
The argument of the first delta functional vanishes only when
$D\!*\!D\!*\!u_a=0$; assuming that the connection $\omega$ is irreducible,
this implies that $u_a=0$.  The delta functional then imposes the condition
\beq
{\bar D}\Omega_a + {1\over2}\epsilon_{abc}\,\Omega^b\wedge\Omega^c = 0 ,
\label{c8}
\eeq
which can be recognized as the requirement that $\omega = \bar\omega +
\Omega$ be a flat connection.  This, in turn, allows us to eliminate the
term
$$\int_M {1\over2}\epsilon_{abc}\,{\bar e}^a\wedge\Omega^b\wedge\Omega^c =
 -\int_M {\bar e}^a\wedge{\bar D}\Omega_a
 = \int_M {\bar D}{\bar e}^a\wedge\Omega_a = 0$$
in \rref{c6}.

We can now use the delta functionals to perform the remaining integration
over $\Omega$.  It is straightforward to show that
\beq
[d\Omega]
\delta\left[{\bar D}\Omega_a
  + {1\over2}\epsilon_{abc}\,\Omega^b\wedge\Omega^c
  + *D\!*\!u_a\right] \delta\left[*D\!*\!\Omega_a\right]
  = [d\tilde\omega] |\hbox{det}'\tilde L_-^{\hbox{\scriptsize rel}}|^{-1} ,
\label{c9}
\eeq
where $\tilde\omega=\bar\omega+\tilde\Omega$ ranges over flat connections
with our specified boundary values and $\tilde L_-^{\hbox{\scriptsize rel}}
= *D_{\tilde\omega} + D_{\tilde\omega}*$ maps a one-form plus a three-form
$(\alpha,\beta)$ obeying relative boundary conditions to a one-form plus
a three-form
$(*D_{\tilde\omega}\alpha+D_{\tilde\omega}*\beta, D_{\tilde\omega}*\alpha)$
obeying absolute boundary conditions.  Performing the ghost integrals, we
finally obtain
\beq
Z_M[I^*\omega] = \int [d\tilde\omega][d\tilde E]
  {\hbox{det}'\tilde\Delta_{(0)}^{\hbox{\scriptsize rel}}
  \hbox{det}'\tilde\Delta_{(0)}^{\hbox{\scriptsize abs}}\over
  |\hbox{det}'\tilde L_-^{\hbox{\scriptsize rel}}|} ,
\label{c10}
\eeq
where $\tilde\Delta_{(k)}$ is the Laplacian $*D_{\tilde\omega}
\!*\!D_{\tilde\omega} + D_{\tilde\omega}\!*\!D_{\tilde\omega}*$
acting on $k$-forms.

(Strictly speaking, the determinants $\hbox{det}'\tilde\Delta$ are not
well-defined for a noncompact gauge group like SO(2,1).  But this is a
minor problem, whose solution has been discussed in \cite{BarNatan} and
\cite{Muller2}, and it does not affect our final expression for amplitudes
in terms of Reidemeister torsion \cite{Muller2}.)

Now, by expanding one-forms and three-forms in modes of $L_-^\dagger L_-$,
one may easily show that
\beq
|(\hbox{det}'\tilde L_- )(\hbox{det}'\tilde L_-^\dagger)|
 = \hbox{det}'\tilde\Delta_{(1)}^{\hbox{\scriptsize rel}}
   \hbox{det}'\tilde\Delta_{(3)}^{\hbox{\scriptsize rel}}
 = \hbox{det}'\tilde\Delta_{(1)}^{\hbox{\scriptsize abs}}
   \hbox{det}'\tilde\Delta_{(3)}^{\hbox{\scriptsize abs}}
\label{c11}
\eeq
Moreover, $\hbox{det}'\tilde\Delta_{(k)}^{\hbox{\scriptsize rel}} =
\hbox{det}'\tilde\Delta_{(3-k)}^{\hbox{\scriptsize abs}}$, since
the Hodge star operator maps any eigenfunction $\alpha$ of
$\tilde\Delta_{(k)}^{\hbox{\scriptsize rel}}$ to an eigenfunction
$*\alpha$ of $\tilde\Delta_{(3-k)}^{\hbox{\scriptsize abs}}$ with
the same eigenvalue.  Some simple manipulation then shows that
\begin{eqnarray}
&&Z_M[I^*\omega] = \int [d\tilde\omega][d\tilde E] T[\tilde\omega] ,
\nonumber\\
&&T[\tilde\omega] =
{(\hbox{det}'\tilde\Delta_{(3)}^{\hbox{\scriptsize rel}})^{3/2}
(\hbox{det}'\tilde\Delta_{(1)}^{\hbox{\scriptsize rel}})^{1/2}\over
(\hbox{det}'\tilde\Delta_{(2)}^{\hbox{\scriptsize rel}})} =
{(\hbox{det}'\tilde\Delta_{(3)}^{\hbox{\scriptsize abs}})^{3/2}
(\hbox{det}'\tilde\Delta_{(1)}^{\hbox{\scriptsize abs}})^{1/2}\over
(\hbox{det}'\tilde\Delta_{(2)}^{\hbox{\scriptsize abs}})}.
\label{c13}
\end{eqnarray}
The combination of determinants $T[\tilde\omega]$ may be recognized as
the Ray-Singer torsion \cite{RaySinger}; the equality of relative and
absolute torsions on odd-dimensional manifolds is shown in \cite{Luck}.
A similar transition amplitude for Abelian $B$-$F$ theories has bee
discussed by Wu \cite{Wu}.

We can now return to the question of whether our amplitude $Z_M$
depends on the choice of auxiliary metric $h$.  When there are no
zero-modes, $T[\tilde\omega]$ is known to be independent of $h$
\cite{RaySinger,Lott}.  When zero-modes are present, $T[\tilde\omega]$
depends on $h$, but so does the volume element $[d\tilde\omega]
[d\tilde E]$.  If, as we have assumed, there are no ghost
zero-modes---that is, if $H^0(M;V_{\tilde\omega}) = H^0(M,\partial M;
V_{\tilde\omega}) = 0$---then the combination $[d\tilde\omega]
[d\tilde E]T[\tilde\omega]$ is again independent of $h$
\cite{Roth,Vishik}.  If ghost zero-modes are present, they must
be included in the integral \rref{c13}; when they are, the amplitude
is once again independent of $h$.

We conclude this section with a discussion of the range of integration
in \rref{c13}.  As noted above, $\tilde\omega$ ranges over the space of
gauge-fixed flat SO(2,1) connections---i.e., the moduli space of flat
connections modulo gauge transformations---with specified boundary
values.  In general, this space is rather complicated.\footnote{For
the remainder of this paper, we shall treat this space as if it were
a manifold.  When $M$ has the topology $\IR\!\times\!\Sigma$, this is
essentially correct \cite{Goldman}; for more complicated topologies,
we do not know whether this assumption is justified.}  We can at least
determine its dimension, however, by linearizing \rref{c3} and
\rref{c8}: solutions $\delta\Omega$ of
\beq
D_{\tilde\omega} {\delta\Omega^a} = 0 = *D_{\tilde\omega}*{\delta\Omega^a}
\label{c14}
\eeq
satisfying the boundary conditions \rref{b8} are cotangent vectors to the
moduli space of flat connections at $\tilde\omega$.

These conditions have a natural cohomological interpretation.  Since the
connection $\tilde\omega$ is flat, $D_{\tilde\omega}^2=0$, so we can
construct a de Rham cohomology $H^*(M;V_{\tilde\omega})$ on the complex
of forms on $M$ with values in the flat bundle $V_{\tilde\omega}$
determined by $\tilde\omega$.  Then \rref{c14} is equivalent to the
condition
\beq
\delta\Omega \in H^1(M,\partial M;V_{\tilde\omega}) ,
\label{c15}
\eeq
where the boundary conditions determine the use of relative cohomology.
The range of $\tilde E$ has a similar interpretation.  From \rref{c3}
and \rref{c6a} and the boundary conditions \rref{b10}, it follows that
\beq
\tilde E \in H^1(M;V_{\tilde\omega}) .
\label{c16}
\eeq
Because of the change in boundary conditions, $\tilde E$ and $\delta\Omega$
do not lie in the same cohomology groups; indeed, by Poincar{\'e}-Lefschetz
duality,
\beq
H^1(M;V_{\tilde\omega}) \approx H^2(M,\partial M;V_{\tilde\omega}) .
\label{c17}
\eeq

Unlike \rref{c15}, equation \rref{c16} determines the entire space of
fields $\tilde E$, not merely its tangent space.  Moreover, the integrand
$T[\tilde\omega]$ in \rref{c13} is independent of $\tilde E$.  This means
that if the cohomology group \rref{c17} is nontrivial, the amplitude
$Z_M[I^*\omega]$ diverges.  This is the infrared divergence cited by
Witten as an indication of classical behavior \cite{Wit2}.  In contrast
to the cases discussed by Witten, however, our boundary conditions allow
this divergence to appear even when the moduli space of flat connections
$\tilde\omega$ consists of isolated points.  This will be the case in
specific examples we discuss below.

\section{From Ray-Singer to Reidemeister Torsion}
\addtocounter{footnote}{-3}

In principle, the integral \rref{c13} determines the transition amplitude
for an arbitrary topology change in 2+1 dimensions.  In practice, however,
the determinants in $T[\tilde\omega]$ are usually impossible to evaluate.
We must therefore take one further step, and relate the Ray-Singer torsion
to the combinatorial, or Reidemeister, torsion.

We begin with a brief description of the Reidemeister torsion.  (For more
details, see \cite{RaySinger}, \cite{Milnor}, or \cite{Muller}).  It is
instructive to start with a concrete description of the flat bundle
$V_{\tilde\omega}$.  Let $\pi_1$ be the fundamental group of $M$,
and let $\widetilde M$ denote the universal covering space of $M$, so
\beq
M \approx \widetilde M/\pi_1 .
\label{d1}
\eeq
As in \rref{b6}, the flat connection $\tilde\omega$ is determined up to
gauge transformations by its holonomy group
\beq
H: \pi_1 \rightarrow \hbox{SO(2,1)} .
\label{d2}
\eeq
We can now define our flat bundle as
\beq
V_{\tilde\omega}
 = \left( \widetilde M\!\times\!\hbox{\sl so}(2,1) \right)/\pi_1,
\label{d3}
\eeq
where $\pi_1$ acts on $\widetilde M$ as in \rref{d1} and on the Lie
algebra $\hbox{\sl so}(2,1)$ by the adjoint action of the holonomy group.

We now repeat this construction in a slightly different form.  Let
us treat $M$ as a CW complex, with $k$-cells $\{e_{(k)}^\alpha\}$.
$M$ can be viewed as a fundamental domain embedded in $\widetilde M$,
which then has a corresponding cell decomposition in terms of the
translates $\{ge_{(k)}^\alpha ,\, g\!\in\!\pi_1\}$.  The chain groups
$C_k(\widetilde M)$ of the universal covering space $\widetilde M$ thus
become modules over the real group algebra $\IR[\pi_1]$, with the
$\{e_{(k)}^\alpha\}$ constituting a preferred basis.\footnote{As usual,
we denote chain groups by $C_k$, the kernel of $\partial$ in $C_k$ by
$Z_k$, and the image of $C_{k+1}$ in $Z_k$ by $B_k$; the homology groups
are $H_k = Z_k/B_k$.}  Note that if $e_{(k)}^\alpha$ is a cell in $M$,
its boundary will not, in general, lie in $M$, but will rather be a sum
of translates of $(k-1)$-cells $e_{(k)}^\beta$ by elements of $\pi_1$.
Relative to our preferred basis, the boundary operator can thus be viewed
as a matrix with elements in $\IR[\pi_1]$.

We can now construct the twisted chain complex
\beq
{\cal C}(\widetilde M; V_{\tilde\omega}) =
  {\cal C}(\widetilde M)\otimes_{R[\pi_1]}\hbox{\sl so}(2,1) ,
\label{d4}
\eeq
where, as in \rref{d3},  $\pi_1$ acts on ${\cal C}(\widetilde M)$ by
translation and on $\hbox{\sl so}(2,1)$ by the adjoint action.  A
preferred basis for ${\cal C}(\widetilde M; V_{\tilde\omega})$ consists
of the elements $\{e_{(k)}^\alpha\otimes J^a\}$, where the $J^a$ are
an orthonormal set of generators of $\hbox{\sl so}(2,1)$.  We shall
abbreviate these basis elements by ${\bf c}_{(k)}$ below.  The boundary
operator for this twisted complex can be viewed as a real matrix: if
$\partial e_{(k)}^\alpha = ge_{(k-1)}^\beta + \dots$, then
\beq
\partial\, (e_{(k)}^\alpha\otimes J^a)
  = e_{(k-1)}^\beta \otimes ad(g)J^a + \dots
  = e_{(k-1)}^\beta \otimes S[g]^a{}_b J^b + \dots ,
\label{d5}
\eeq
where $S[g]^a{}_b$ are the matrix elements of $g$ in the adjoint
representation.

The Reidemeister torsion of $M$ is now defined as follows.  We have
chosen a preferred basis ${\bf c}_{(k)}$ for each of the chain groups
$C_k(\widetilde M; V_{\tilde\omega})$.  We also have a preferred basis
for each homology group $H_k(\widetilde M; V_{\tilde\omega})$, determined
from the harmonic forms of the previous section by the de Rham isomorphism.
Let us select a set ${\bf\tilde h}_{(k)}\in Z_k$ to represent these
basis elements.  We next choose an arbitrary basis ${\bf b}_{(k)}$ for
each $B_k$, and a set ${\bf \tilde b}_{(k-1)}$ in $C_k$ such that
$\partial{\bf \tilde b}_{(k-1)} = {\bf b}_{(k-1)}$.

It is easy to see that the set $({\bf b}_{(k)},{\bf\tilde h}_{(k)},
{\bf\tilde b}_{(k-1)})$ forms a new basis---call it
${\bf\hat c}_{(k)}$---for $C_k$.  Let $T_{(k)}$ denote the
matrix representing the change of basis from ${\bf c}_{(k)}$ to
${\bf\hat c}_{(k)}$, that is, ${\bf\hat c}_{(k)} = T_{(k)}{\bf c}_{(k)}$.
The Reidemeister torsion is then defined as
\beq
\tau(M;V_{\tilde\omega}) = {\hbox{det}\,T_{(0)}\hbox{det}\,T_{(2)}\over
  \hbox{det}\,T_{(1)}\hbox{det}\,T_{(3)}} .
\label{d6}
\eeq
The relative Reidemeister torsion $\tau(M,\partial M; V_{\tilde\omega})$
is obtained by the same construction with the relative chain complex.
Since cellular decompositions of manifolds are often rather simple, this
invariant can sometimes be calculated quite explicitly.

For a manifold without boundary, the Reidemeister torsion is, remarkably,
equal to the Ray-Singer torsion $T[\tilde\omega]$ of equation \rref{c13}
\cite{Muller,Cheeger}.  When a boundary is present, this equality no
longer holds, but only a small correction is needed \cite{Luck}:
\beq
T[\tilde\omega]
 = 2^{3\chi(\partial M)/4}\,\tau(M,\partial M; V_{\tilde\omega}) ,
\label{d7}
\eeq
where $\chi(\partial M)$ is the Euler characteristic of the boundary.
(The factor of three in the exponent is the dimension of {\sl so}(2,1).)
We can therefore determine the transition amplitude \rref{c13} by
evaluating the Reidemeister torsion for $M$.

\section{Topology-Changing Manifolds}

To compute a topology-changing amplitude, it remains for us to specify
the manifold $M$ that mediates the transition.  This choice is not
trivial---as Amano and Higuchi have shown \cite{Amano}, the requirement
that the initial and final boundaries be spacelike strongly restricts the
allowed topologies.  We begin with $M_2$, the simplest of the interpolating
manifolds of reference \cite{Amano}, which describes a transition from a
genus three surface to two genus two surfaces.  This manifold, shown in
figure 1, can be described as follows:

Let $V$ be a genus four handle-body, and let $\bar V$ denote a reflected
copy of $V$.  Remove from the interior of $V$ two genus two handle-bodies,
as indicated by the shaded areas in figure 1, to obtain a manifold $W_1$
with boundary $\partial W_1\approx\partial V\cup\Sigma_2\cup\Sigma_2'$,
where $\Sigma_2$ and $\Sigma_2'$ are genus two surfaces.  Next, remove
from the interior of $\bar V$ a single genus three handle-body, as shown,
to obtain a manifold $W_2$ with boundary $\partial W_2\approx\partial
{\bar V}\cup\Sigma_3$, where $\Sigma_3$ is a genus three surface.  Now
identify $W_1$ and $W_2$ along their common boundary $\partial V\approx
\partial{\bar V}$ to obtain our desired manifold $M_2$, which clearly
has a boundary $\partial M_2\approx\Sigma_2\cup\Sigma_2'\cup\Sigma_3$.
More general amplitudes can be obtained by combining manifolds $M_n$
that mediate between a genus $n+1$ surface and $n$ genus two surfaces;
we refer the reader to reference \cite{Amano} for details.

Figure 1 also indicates a set of generators $\rho_1,\dots,\rho_8$ for
$\pi_1(M_2)$, which obey the relations
\begin{eqnarray}
&&[\rho_2^\zx,\rho_1^{-1}][\rho_3^{-1},\rho_4^{-1}] =
  [\rho_6^\zx,\rho_5^{-1}][\rho_7^{-1},\rho_8^{-1}] = 1 ,\nonumber\\
&&\rho_3=\rho_5 ,
\label{e1}
\end{eqnarray}
where $[\sigma,\tau] = \sigma^\zx\tau^\zx\sigma^{-1}\tau^{-1}$ is the
commutator.  Apart from a slightly unusual normalization that will be
useful later in describing the cell decomposition, the relation obeyed
by $\rho_1,\dots,\rho_4$ can be recognized as that of a Fuchsian group
that uniformizes a genus two surface \cite{Abikoff}.  The generators
$\rho_5,\dots,\rho_8$ determine a similar group.

It will be helpful to compare the amplitude coming from $M_2$ to a
related topology-preserving amplitude.  We obtain the latter from a new
manifold $P_2$ constructed by attaching two copies of $W_2$ along their
common boundary $\partial V$.  This manifold mediates transitions from
$\Sigma_3$ to another genus three surface $\Sigma_3'$.  Figure 2 shows
$P_2$, along with a set of generators $\sigma_1,\dots,\sigma_8$ of
$\pi_1(P_2)$, which satisfy the relations
\beq
\sigma^\zx_4[\sigma^\zx_2,\sigma_1^{-1}]
  [\sigma_3^{-1},\sigma_4^{-1}]\sigma_4^{-1} =
\sigma_6^{-1}[\sigma_8^{-1},\sigma_7^{-1}]
  [\sigma_3^{-1},\sigma^\zx_6]\sigma^\zx_6 = \sigma^\zx_5 .
\label{e2}
\eeq

To compute the Reidemeister torsion, we need a cell decomposition for the
universal covering space $\widetilde M_2$.  The first step in obtaining
such a decomposition is to dissect $M_2$ into a simply connected region
$U_M$ that can serve as a fundamental domain upon which $\pi_1(M_2)$ acts
by translation.  We do so by cutting $M_2$ open along a set of surfaces
transverse to the generators of the fundamental group.

Since this process is somewhat difficult to visualize, let us begin with
a simple example.  Let $\Sigma_{1,1}$ be a genus one surface with a single
hole, and consider a thickening of $\Sigma_{1,1}$ into a three-manifold
$N$, one of the elementary building blocks of $M_2$.  Figure 3 shows a
dissection of $N$ into a simply connected fundamental domain.  It is clear
that $N$ can be recovered by identifying the boundaries of this figure;
this identification represents the action of $\pi_1(N)$.

Figure 4 shows the corresponding dissection of $M_2$.  The cuts on the
top half of the figure are evidently analogous to those of figure 3;
they are then extended through the boundary $\partial V$ into the bottom
half of the figure, where they are continued until they reach the inner
boundary $\Sigma_3$.

The corresponding construction for $P_2$ is illustrated in figure 5.
We start with two copies of the bottom half of figure 4.  However, a new
noncontractible loop now appears, representing the element $\sigma_5$ of
$\pi_1(P_2)$, that was not present in figure 4.  To obtain a simply
connected fundamental domain $U_P$ for $P_2$, we therefore need one more
cut transverse to this loop.  This final cut, required by the larger
fundamental group of $P_2$, will play a critical role in the suppression
of topology change---it will lead to the existence of extra zero-modes,
and thus more highly divergent infrared behavior, in the topology-preserving
amplitude.

To obtain cell decompositions of $\widetilde M_2$ and $\widetilde P_2$,
it now suffices to find cell decompositions of $U_M$ and $U_P$
compatible with the action of their fundamental groups---that is,
decompositions for which the process of gluing (say) $U_M$ back together
to obtain $M_2$ identifies like cells.  The final results are shown in
figures 4 and 5.  We have labeled a basis $e_{(k)}^\alpha$ of $k$-cells
in each figure; the remaining cells are translates of these basis
elements by elements of the appropriate fundamental group.

\section{An Explicit Computation}

We are finally ready to calculate the amplitude $Z_{M_2}$ for tunneling
from the genus three surface $\Sigma_3$ to the genus two surfaces
$\Sigma_2$ and $\Sigma_2'$.  We shall proceed in two steps, first
obtaining some general information about the $\tilde\omega$ and
$\tilde E$ integrals in \rref{c13} and then computing the Reidemeister
torsion for a particular choice of flat connection.

We begin by showing that the topology-preserving path integral $Z_{P_2}$
has at least six more zero-modes than $Z_{M_2}$.  This is in itself
enough to indicate a suppression of topology change---each $\tilde E$ mode
leads to an infrared divergence, so this mismatch shows that$Z_{P_2}$
is more divergent than $Z_{M_2}$.  We caution the reader, however, that
the choice of $P_2$ to ``normalize'' $M_2$ is somewhat arbitrary; until
we understand more about the overall normalization of amplitudes, we can
draw no firm conclusions about absolute probabilities for topology change.

Observe first that the twisted Euler characteristics \cite{Maunder}
\begin{eqnarray}
\sum(-1)^i\hbox{\em dim}H_i(M_2,\partial M_2;V_{\tilde\omega})
  &=& \chi(M_2,\partial M_2;V_{\tilde\omega})
   = {1\over2}\chi(\partial M_2;V_{\tilde\omega}) = 12\nonumber\\
\sum(-1)^i\hbox{\em dim}H_i(P_2,\partial P_2;V_{\tilde\omega})
  &=& \chi(P_2,\partial P_2;V_{\tilde\omega})
   = {1\over2}\chi(\partial P_2;V_{\tilde\omega}) = 12
\label{f1}
\end{eqnarray}
serve as constraints on the dimensions of the twisted homology groups,
placing a lower limit of twelve on the total number of zero-modes.  In the
case of $M_2$, this limit is realized.  Indeed, the twisted cell complex
associated with $M_2$ is
\beq
\begin{array}{ccccccc}
C_0(M_2,\partial M_2;V_{\tilde\omega})&\stackrel{\partial_1}{\leftarrow}&
C_1(M_2,\partial M_2;V_{\tilde\omega})&\stackrel{\partial_2}{\leftarrow}&
C_2(M_2,\partial M_2;V_{\tilde\omega})&\stackrel{\partial_3}{\leftarrow}&
C_3(M_2,\partial M_2;V_{\tilde\omega}) . \\
\|&&\|&&\| &&\|\\
0 &\stackrel{\partial_1}{\leftarrow}&
  \underbrace{\IR\oplus\dots\oplus\IR}_{6\,{\hbox{\scriptsize times}}}
  &\stackrel{\partial_2}{\leftarrow}&
  \underbrace{\IR\oplus\dots\oplus\IR}_{21\,{\hbox{\scriptsize times}}}
  &\stackrel{\partial_3}{\leftarrow}& \IR\oplus\IR\oplus\IR
\end{array}
\label{f2}
\eeq
If we can show that $\partial_2$ is an epimorphism and that $\partial_3$ is
a monomorphism, it will follow that $H_0(M_2,\partial M_2;V_{\tilde\omega})
= H_1(M_2,\partial M_2;V_{\tilde\omega}) = H_3(M_2,\partial M_2;
V_{\tilde\omega}) = 0$, and hence that ${\em dim}H_2(M_2,\partial M_2;
V_{\tilde\omega}) = 12$.

Given a set of generators $\{J^0,J^1,J^2\}$ for SO(2,1), the adjoint action
described in section 4 determines a matrix representation of $\pi_1(M_2)$,
$$S\!: \pi_1(M_2)\rightarrow\hbox{GL($3,\IR$)} .$$
To show that $\partial_2$ is an epimorphism, consider the following six
boundary operations, which can be read off figure 4:
\begin{eqnarray}
\partial_2(e^1_{(2)}\otimes J^a) &=& e^1_{(1)}\otimes(1-S[\rho^\zx_3]) J^a
  \nonumber\\
\partial_2(e^3_{(2)}\otimes J^a) &=&
  e^1_{(1)}\otimes S[\rho^\zx_2](1-S[\rho^\zx_1])
  S[\rho_3^{-1}\rho_4^{-1}\rho^\zx_3] J^a .
\label{f3}
\end{eqnarray}
The $S[\rho_i]$ are nondegenerate matrices which each stabilize a
one-dimensional subspace of $\IR^3$, so $1-S[\rho_3]$ is a rank two
matrix that is zero on the subspace stabilized by $S[\rho_3]$.
Similarly, $S[\rho^\zx_2](1-S[\rho^\zx_1])S[\rho_3^{-1}\rho_4^{-1}
\rho^\zx_3]$ is a rank two matrix that is zero on the subspace
stabilized by $S' = S[\rho_3^{-1}\rho_4^\zx\rho^\zx_3\rho^\zx_1
\rho_3^{-1}\rho_4^{-1}\rho^\zx_3]$.  But in order for the boundary
$\partial M_2$ to be spacelike, the commutator
$[\rho_3^\zx,
\rho_3^{-1}\rho_4^\zx\rho^\zx_3\rho^\zx_1\rho_3^{-1}\rho_4^{-1}
\rho^\zx_3]$,
which represents a loop on $\partial M_2$, must be nontrivial.  Hence
$S[\rho_3]$ and $S'$ stabilize different one-dimensional subspaces,
and the images of $1-S[\rho_3]$ and $1-S'$ span all of $\IR^3$.  This,
in turn, means that the image under $\partial_2$ of the space spanned by
$\{e^1_{(2)}\otimes J^a, e^3_{(2)}\otimes J^a\}$ contains $\{e^1_{(1)}
\otimes\hbox{\sl so}(2,1)\}$.  A similar argument shows that the image
of the space spanned by $\{e^2_{(2)}\otimes J^a, e^4_{(2)}\otimes J^a\}$
contains $\{e^2_{(1)}\otimes\hbox{\sl so}(2,1)\}$, thus establishing that
$\partial_2$ is an epimorphism.

An analogous argument, starting with the boundary operations
\beq
\partial_3(e_{(3)}\otimes J^a)  = e_{(2)}^1\otimes(1-S[\rho_4^{-1}]) J^a
  - e_{(2)}^7\otimes(1-S[\rho_3]) J^a + \dots
\label{f4}
\eeq
and the nontriviality of the commutator $[\rho_4^{-1},\rho_3^\zx]$,
shows that $\partial_3$ is a monomorphism.  We have thus established
that $H_2(M_2,\partial M_2;V_{\tilde\omega})$ is twelve-dimensional.
This conclusion has been checked by making  explicit choices for the
holonomies $\{S[\rho_i]\}$ and identifying generators of the twisted
homology groups.

We can repeat the same analysis for $P_2$, but let us instead take a
shortcut.  It is easy to see that $H_0(P_2,\partial P_2;V_{\tilde\omega})
= 0$, so the constraint \rref{f1} implies that
$$\hbox{\em dim}H_2(P_2,\partial P_2;V_{\tilde\omega})\ge 12 +
 {\em dim}H_1(P_2,\partial P_2;V_{\tilde\omega}).$$
So if we can show the existence of three generators of $H_1(P_2,\partial
P_2;V_{\tilde\omega})$, we will have proven that $Z_{P_2}$ has at least
six more zero-mode integrations---three $\tilde\omega$ modes and three
$\tilde E$ modes---than $Z_{M_2}$.

But this is apparent from figure 5: no combination of $e^3_{(1)}\otimes
J^a$ and $e^4_{(1)}\otimes J^a$ lies in the image of $\partial_2$, and
\beq
\partial_1\left( e^3_{(1)}\otimes J^a + e^4_{(1)}\otimes J^a \right) = 0 .
\label{f5}
\eeq
Hence $\{ (e^3_{(1)} + e^4_{(1)})\otimes J^a\}$ represent three generators
of $H_1(P_2,\partial P_2; V_{\tilde\omega})$, as claimed.

We conclude this section by describing an explicit computation of
the Reidemeister torsion for $M_2$.  This requires a choice of flat
connection $\tilde\omega$, which may be determined by its holonomies
around the curves $\rho_i$ that generate $\pi_1(M_2)$.  Note, however,
that each of the $\rho_i$ can be deformed to a curve on one of the
boundary components of $M_2$.  The connection $\tilde\omega$ is therefore
fixed by its holonomies on $\Sigma_2\cup\Sigma_2'\cup\Sigma_3$, and hence
by the boundary data $I^*\omega$.  This means that the relevant moduli
space of flat connections is a single point, and that the integral
over $\tilde\omega$ disappears from \rref{c13}, as expected from the
vanishing of $H_1(M_2,\partial M_2;V_{\tilde\omega})$.

Following reference \cite{Amano}, we choose the $S[\rho_i]$ as follows.
$S[\rho_1]$ through $S[\rho_4]$ are the generators of an arbitrary Fuchsian
group uniformizing a genus two surface.  Such groups form a six-parameter
family, which can be written down from Fenchel-Nielsen coordinates by
using the results of \cite{Wolpert}.  A convenient two-parameter family
$S[\rho_i](k,r)$ is given in the appendix.  We then take
\begin{eqnarray}
S[\rho_5^\zx] &=& S[\rho_3^\zx] ,
 \quad S[\rho_6^\zx] = S[\rho_4^\zx] , \nonumber\\
S[\rho_7^\zx] &=& S[\rho_4^\zx\rho_2^\zx\rho_1^\zx\rho_2^{-1}\rho_4^{-1}] ,
  \quad S[\rho_8^\zx] = S[\rho_4^\zx\rho_2^\zx\rho_4^{-1}] .
\label{f7}
\end{eqnarray}
This expression differs slightly from that of reference \cite{Amano},
but only because of our different choice of generators of $\pi_1(M_2)$.
Amano and Higuchi show that with these choices, the boundaries of $M_2$
are spacelike and nonsingular.

As described in section 4, calculating the torsion requires computing
the determinants of the matrices $T_{(k)}$ that give the change of
basis from ${\bf c}_{(k)}$ to $({\bf b}_{(k)},{\bf\tilde h}_{(k)},
{\bf\tilde b}_{(k-1)})$.  The total measure appearing in the amplitude
\rref{c13} is independent of the choice of basis $({\bf b}_{(k)},
{\bf\tilde h}_{(k)},{\bf\tilde b}_{(k-1)})$, but the determinants
\rref{d6} by themselves are not---we must make an explicit choice of
the homology basis ${\bf\tilde h}_{(k)}$ for a numerical value of the
torsion $\tau$ to have meaning.  For now, we choose the simple but
rather arbitrary basis ${\bf\tilde h}_{(k)}$ described in the appendix.
A tedious but routine calculation then gives a Reidemeister torsion
as a complicated rational function of $k$ and $r$.  A plot is shown
in figure 6.  The torsion falls off very rapidly for large values of
$k$ and $r$; for example,
$$\tau(r=10,k=1000)\approx 7\times10^{-52} .$$

To interpret these results, we must understand the remaining $\tilde E$
integrals in \rref{c13}.  The measure $d\tilde E$ is determined by the
homology basis ${\bf\tilde h}_{(2)}$.  We first select an orthonormal
basis $\tilde E^\alpha$ such that
\beq
\int_{{}^*\tilde h_{(2)}^\alpha} \tilde E^\beta = \delta^\beta_\alpha ,
\label{f9}
\eeq
where ${}^*\tilde h_{(2)}^\alpha$ is the union of one-cells dual to
$\tilde h_{(2)}^\alpha$ and the integral includes a trace over the
generators of {\sl so}(2,1) (see \cite{RaySinger} for details).  If
we then decompose an arbitrary zero-mode as $\tilde E = \sum c_\alpha
\tilde E^\alpha$, the integral \rref{c13} is $\int dc_\alpha$,
and the dependence of $c_\alpha$ on the choice of homology basis
cancels the dependence of the torsion $\tau$.

If we wish to cut off the infrared divergent $\tilde E$ integrals,
however, this basis dependence reappears---the {\em range} of integration
will, in general, depend on the choice of ${\bf\tilde h}_{(2)}$.  This
dependence can be translated into a dependence on the auxiliary metric
$h$ introduced in section 3 to fix the gauge.  The difficulty appears
to be a typical regularization problem---we do not know how to regulate
our integrals in a way that respects the invariance of the original theory.

There is some hope for a physical resolution of this problem.  Witten
has suggested that the infrared divergences of transition amplitudes in
(2+1)-dimensional gravity reflect the appearance of infinite-volume
``classical'' spacetimes.  If we could give a concrete geometrical
meaning to the limits of integration, cutting off at some (observable)
scale, we might be able to define a basis-independent regularization.
It is also possible that the addition of matter to our vacuum theory
might regulate the divergences, again by limiting maximum lengths.  We
leave these questions for future investigation.

\section{Conclusion}

We have now seen that path integrals representing spatial topology
change in (2+1)-dimensional general relativity need not vanish.
Starting with any initial data in our two-parameter family---or,
by a long but straightforward generalization, any other admissible
initial geometry---we can explicitly compute the torsion $T[I^*\omega]$,
and thus the amplitude \rref{c13}.  Indeed, we have seen that such
topology-changing amplitudes may diverge, thanks to the existence
of zero-modes $\tilde E^a$ of the triad $e^a$.  These divergences
presumably reflect the appearance of ``classical'' spacetimes, in
which distances measured with the metric $g_{\mu\nu}=e_\mu{}^ae_{\nu a}$
become arbitrarily large \cite{Wit2}.

Nevertheless, our results may be interpreted as providing evidence
that topology change is suppressed.  We have seen that while the
topology-changing amplitude mediated by $M_2$ is infrared divergent,
the closely related topology-preserving amplitude mediated by $P_2$
is even more divergent.  Evidently, no firm conclusions about topology
change can be drawn without a much better understanding of the overall
normalization of amplitudes in (2+1)-dimensional gravity.  This is a
difficult problem: we must not only consider an infinite number of
possible interpolating manifolds, but must also find a sensible way to
regulate infrared divergences without breaking the symmetries of the
original theory.  Clearly, much work remains to be done.

One important step would be to find an easier and more intuitive method
for computing the degree of divergence for an arbitrary interpolating
manifold without requiring a full cell decomposition.  We do not have
a complete answer to this problem, but we offer the following observations.
As we have seen, $H_1(M,\partial M; V_\omega)$ counts the dimension of
the moduli space of flat connections on $M$ with specified boundary data.
In general, a flat connection is determined up to gauge transformations
by its holonomies, that is, by an assignment of an element of SO(2,1) to
each independent generator of $\pi_1(M)$.  For our manifold $M_2$, it is
evident from figure 1 that every generator of the fundamental group can
be deformed to the boundary; thus, the connection is completely determined
by boundary data, and $\hbox{\em dim}H_1(M_2,\partial M_2;V_\omega) = 0$.
For $P_2$, on the other hand, one independent generator---$\sigma_4$,
for example---cannot be deformed to the boundary, and accounts for our
three generators of $H_1(M_2,\partial M_2;V_\omega)$.  We do not know
whether this argument can be made rigorous when generalized to an
arbitrary three-manifold, but such an extension may be possible.
Similarly, it may be possible to obtain information about $\hbox{\em dim}
H_2(M,\partial M;V_\omega)=\hbox{\em dim}H_1(M;V_\omega)$ by looking at
connections on the double of $M$.  Note that paths connecting boundary
components of $M$ become closed loops on the double, and may contribute
to the dimensions ${\em dim}H_*(M,\partial M;V_\omega)$.  Finally, the
Euler characteristic constraint \rref{f1} holds for any three-manifold,
and places a useful constraint on the number of divergent integrals.

\vspace{1.5ex}
\begin{flushleft}
\large\bf Acknowledgements
\end{flushleft}

We would like to thank Wolfgang L{\"u}ck and Joel Hass for helpful
mathematical advice.  This research was supported in part by National
Science Foundation grant PHY-93-57203 and Department of Energy grant
DE-FG03-91ER40674.

\appendix\section{Computational Details}

In this appendix, we briefly describe some of the intermediate steps
in the computation of the torsion plotted in figure 6.  We begin with
a choice of holonomies, which we give in the two-dimensional
representation of $\hbox{SL}(2,\IR)$:

\begin{eqnarray}
\rho_1 \mapsto \pmatrix{{{5 + {\sqrt{5}}}\over 2}&
   {{-\left( 1 - {\sqrt{5}} \right) \,k\,{r^2}}\over 2}\cr
   {{-3\,\left( 1 + {\sqrt{5}} \right) }\over {k\,{r^2}}}&
   {{-5 + {\sqrt{5}}}\over 2}\cr} , &\qquad&
\rho_2 \mapsto
    \pmatrix{{{\left(1+{\sqrt{5}}\right)\,\left(2+3\,{k^2}\right)}\over
     {2\,k}}&\left( 1 + {k^2} \right) \,{r^2} \cr
   -{{9k^2 + 4}\over {k^2{r^2}}}&
   {{\left( 1 - {\sqrt{5}} \right) \,\left( 2 + 3\,{k^2} \right)}\over
     {2\,k}}\cr} \nonumber\\
\rho_3 \mapsto \pmatrix{ {{\left( 1 + {\sqrt{5}} \right) \,
    \left( 2 + 3\,{k^2} \right) }\over
     {2\,k}}&-9 - {4\over {{k^2}}}\cr
    1 + {k^2}&{{\left( 1 - {\sqrt{5}} \right) \,
       \left( 2 + 3\,{k^2} \right) }\over {2\,k}}\cr} , &\qquad&
\rho_4 \mapsto \pmatrix{ {{-5 + {\sqrt{5}}}\over 2}&
   {{3\,\left( 1 + {\sqrt{5}} \right) }\over k}\cr
   {{\left( 1 - {\sqrt{5}} \right) \,k}\over 2}&
   {{5 + {\sqrt{5}}}\over 2}\cr} .
\label{ap1}
\end{eqnarray}
These describe a genus two surface consisting of two identical one-holed
tori ($\Sigma_{1,1}$ of figure 3) joined with a relative twist; $r$
parametrizes the twist, while $k$ parametrizes the length of a closed
geodesic on each copy of $\Sigma_{1,1}$.

We next describe the choice of homology basis used in the computation of
the torsion $\tau$.  Using the cell structure shown in Figure 4, we begin
by selecting basis elements ${\bf\tilde b}_{(1)}$ and ${\bf\tilde b}_{(2)}$
as defined in section 4:
\begin{eqnarray}
{\tilde b}^1_{(1)} = e^1_{(2)}\otimes J^0 , &\quad&
{\tilde b}^2_{(1)} = e^3_{(2)}\otimes J^0 , \quad
{\tilde b}^3_{(1)} = e^5_{(2)}\otimes J^0 , \nonumber\\
{\tilde b}^4_{(1)} = e^2_{(2)}\otimes J^0 , &\quad&
{\tilde b}^5_{(1)} = e^4_{(2)}\otimes J^0 , \quad
{\tilde b}^6_{(1)} = e^6_{(2)}\otimes J^0 ,
\label{ap2}
\end{eqnarray}
\beq
{\tilde b}^\alpha_{(2)} = e_{(3)}\otimes J^\alpha , \quad \alpha=1,2,3 .
\label{ap3}
\eeq
Our homology basis is then
\begin{eqnarray}
{\tilde h}^1_{(2)} = e^1_{(2)}\otimes J^2
                   + \sum K^1_\alpha{\tilde b}^\alpha_{(1)} , &\quad&
{\tilde h}^2_{(2)} = e^2_{(2)}\otimes J^1
                   + \sum K^2_\alpha{\tilde b}^\alpha_{(1)} ,\nonumber\\
{\tilde h}^3_{(2)} = e^2_{(2)}\otimes J^2
                   + \sum K^3_\alpha{\tilde b}^\alpha_{(1)} , &\quad&
{\tilde h}^4_{(2)} = e^3_{(2)}\otimes J^1
                   + \sum K^4_\alpha{\tilde b}^\alpha_{(1)} ,\nonumber\\
{\tilde h}^5_{(2)} = e^4_{(2)}\otimes J^1
                   + \sum K^5_\alpha{\tilde b}^\alpha_{(1)} , &\quad&
{\tilde h}^6_{(2)} = e^4_{(2)}\otimes J^2
                   + \sum K^6_\alpha{\tilde b}^\alpha_{(1)} ,\nonumber\\
{\tilde h}^7_{(2)} = e^5_{(2)}\otimes J^1
                   + \sum K^7_\alpha{\tilde b}^\alpha_{(1)} , &\quad&
{\tilde h}^8_{(2)} = e^5_{(2)}\otimes J^2
                   + \sum K^8_\alpha{\tilde b}^\alpha_{(1)} ,\nonumber\\
{\tilde h}^9_{(2)} = e^6_{(2)}\otimes J^1
                   + \sum K^9_\alpha{\tilde b}^\alpha_{(1)} , &\quad&
{\tilde h}^{10}_{(2)} = e^6_{(2)}\otimes J^2
                   + \sum K^{10}_\alpha{\tilde b}^\alpha_{(1)} ,\nonumber\\
{\tilde h}^{11}_{(2)} = e^7_{(2)}\otimes J^1
                   + \sum K^{11}_\alpha{\tilde b}^\alpha_{(1)} , &\quad&
{\tilde h}^{12}_{(2)} = e^7_{(2)}\otimes J^2
                   + \sum K^{12}_\alpha{\tilde b}^\alpha_{(1)} ,
\label{ap4}
\end{eqnarray}
where the coefficients $K^i_\alpha$, whose exact values are not needed for
the computation of the torsion $\tau$, are uniquely determined by the
requirement that $\partial h^\alpha_{(2)} = 0$.

Of course, this choice of homology basis depends on the boundary operator
$\partial$, and therefore on the holonomies $\rho_i$.  The choice \rref{ap4}
appears to be valid for generic values of the holonomies, but there are
points at which linear combinations of our $h_{(2)}^\alpha$ lie in the
image of $\partial_3$.  At these points, \rref{ap4} is no longer a valid
basis, and our computed value of the Reidemeister torsion $\tau$ is not
correct---in fact, it appears to go to zero.  This behavior is an artifact
of our basis choice, and does not affect the integral \rref{c13}.

We do not include the final functional form of $\tau(r,k)$---it would
require three pages---but it is available from the authors.

\newpage
\addtolength{\textheight}{.7in}
\pagestyle{empty}
\begin{flushleft}
\small Figure 1.\ The manifold $M_2$ is formed by identifying $W_1$ and
$W_2$ along their common genus four boundary.  A basis of loops for
$\pi_1(M_2)$ is shown.
\end{flushleft}
\epsfxsize=6in\epsfbox{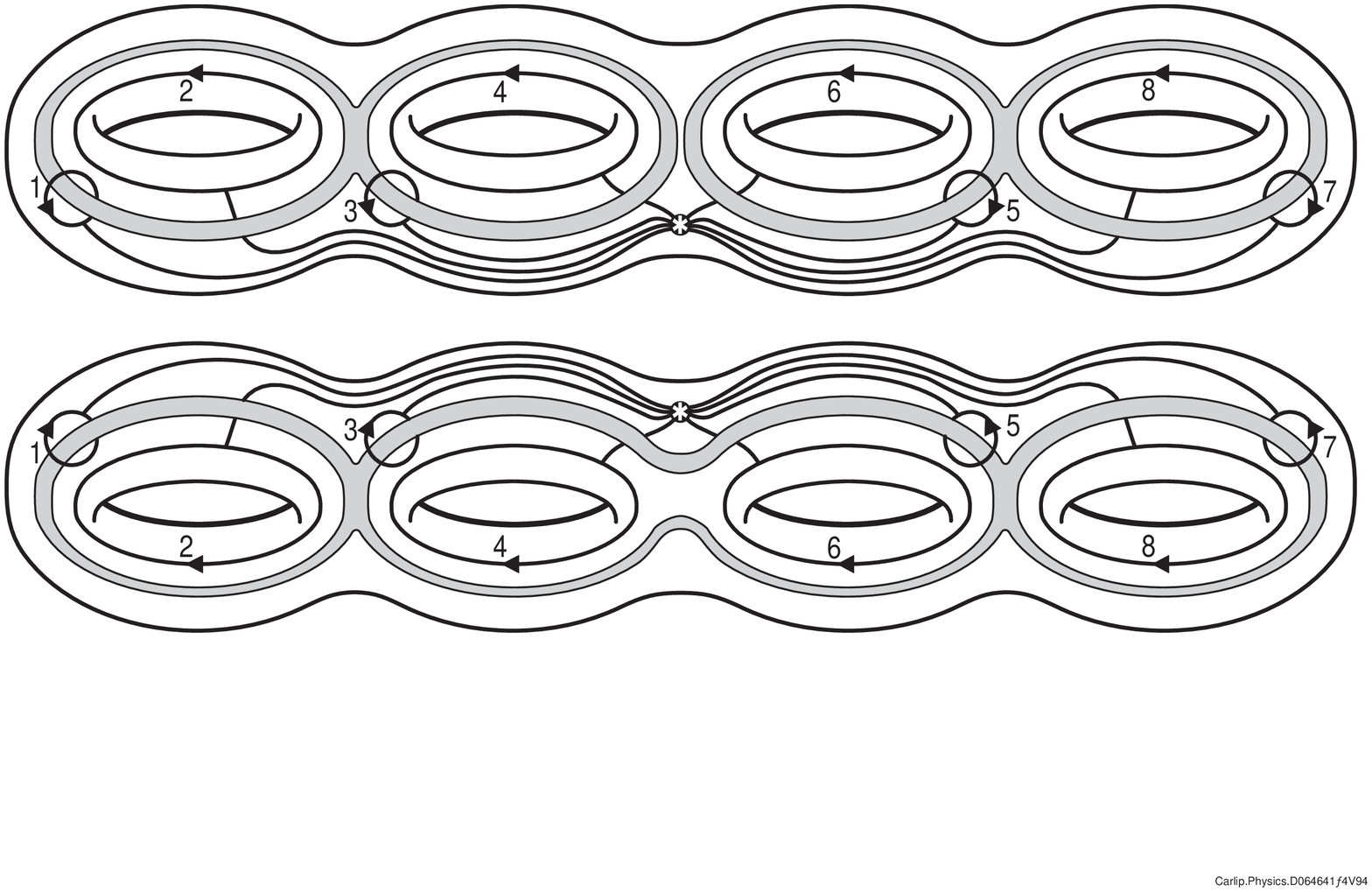}
\begin{flushleft}
\small Figure 2.\ The manifold $P_2$ and a basis for its fundamental group.
\end{flushleft}
\epsfxsize=6in\epsfbox{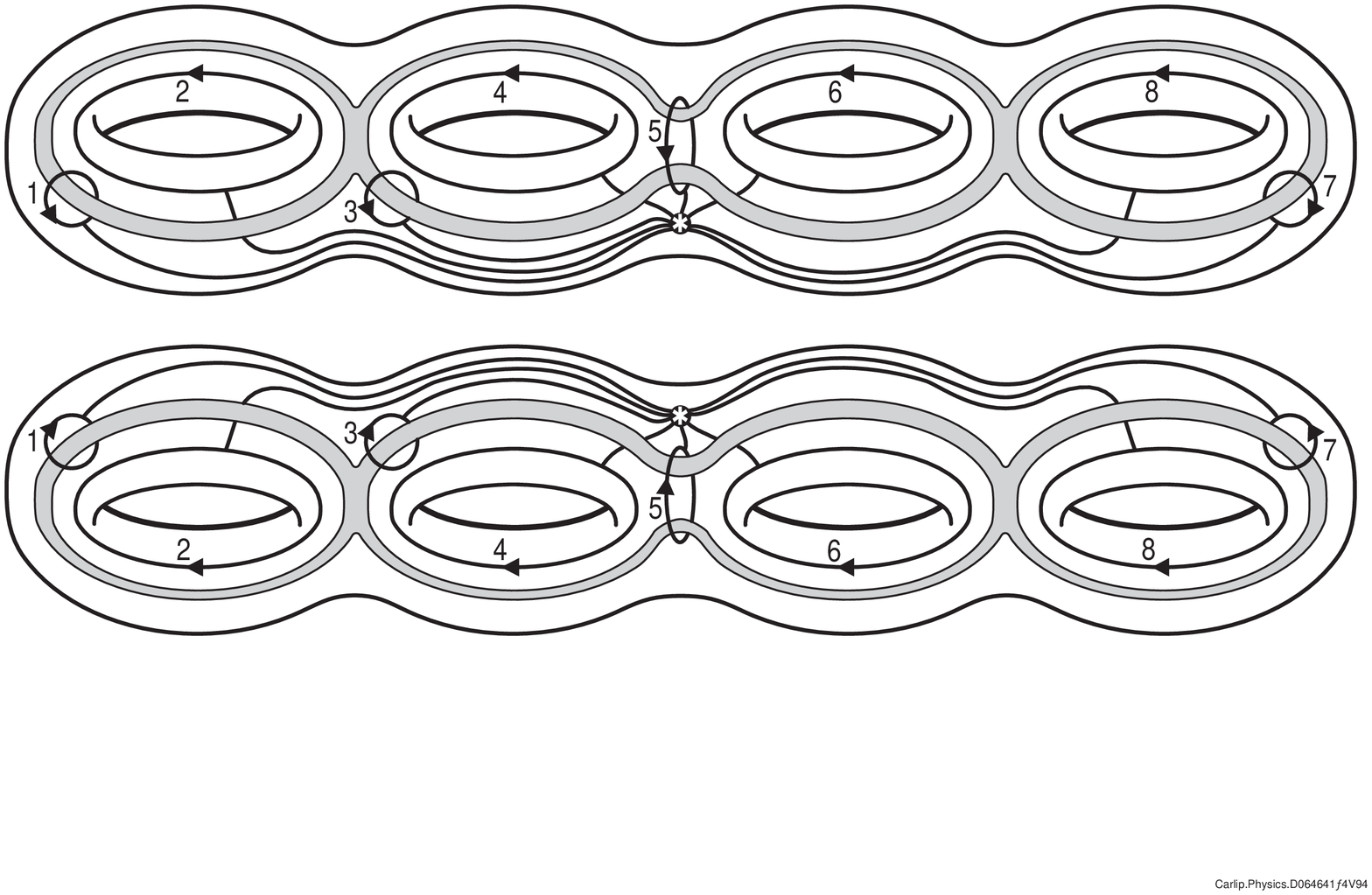}
\newpage
\begin{flushleft}
\small Figure 3.\ A dissection of the thickened one-holed torus into a
simply connected fundamental domain.
\end{flushleft}
\epsfysize=7in\epsfbox{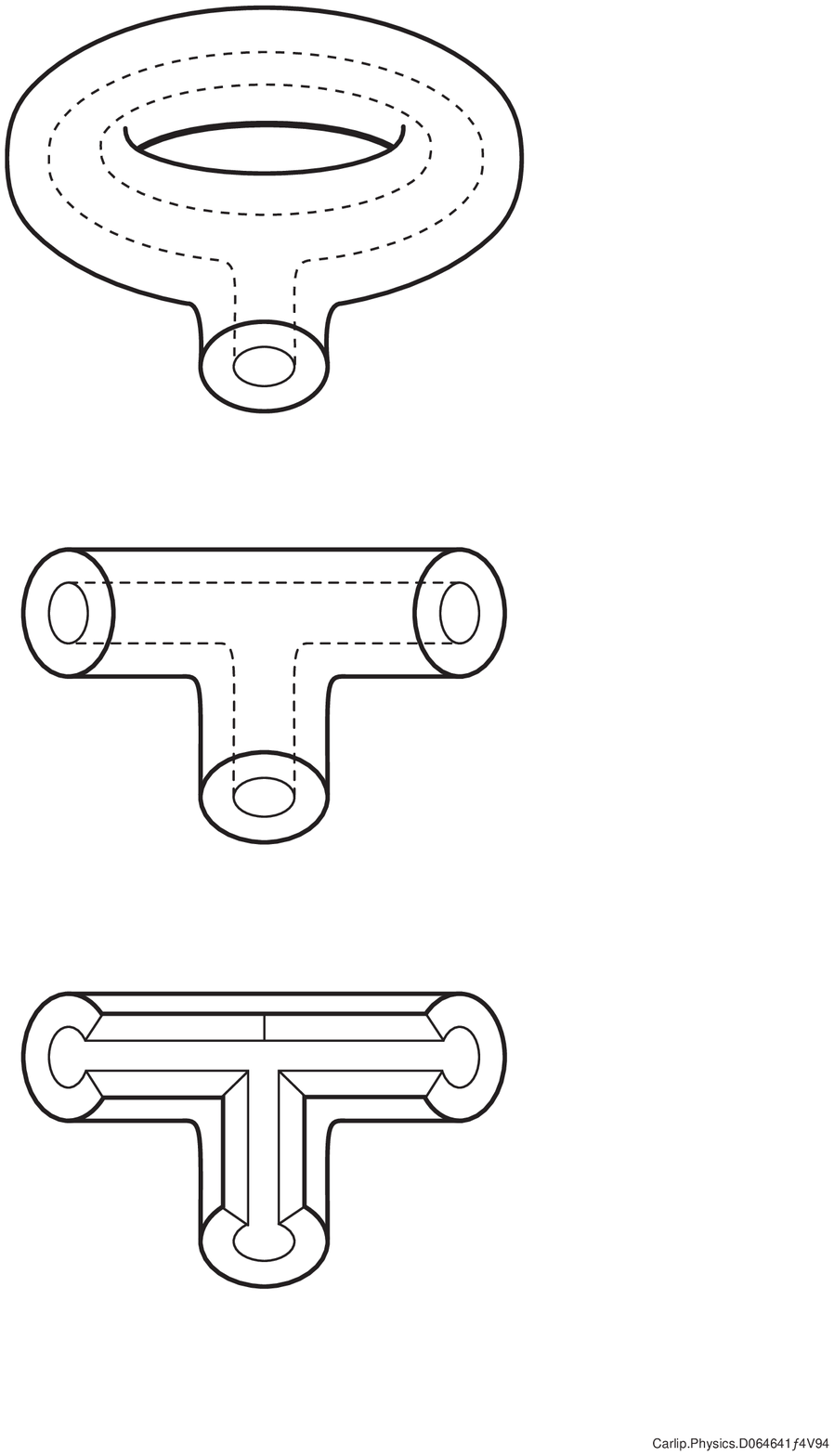}
\newpage
\begin{flushleft}
\small Figure 4.\ A dissection of $M_2$.  A basis of cells $e^\alpha_{(k)}$
is shown.
\end{flushleft}
\epsfxsize=5.7in\epsfbox{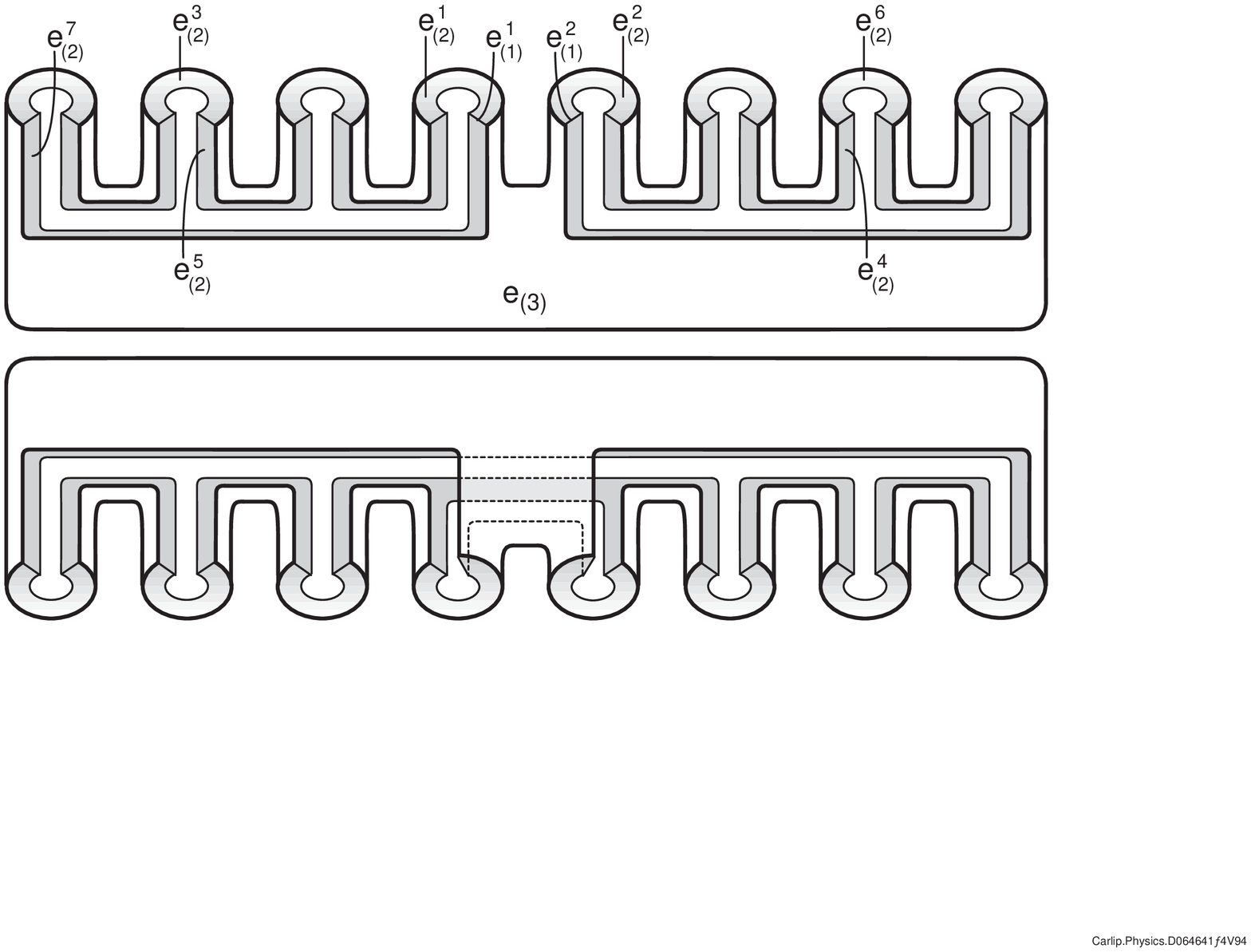}
\begin{flushleft}
\small Figure 5.\ A dissection of $P_2$ and a basis of cells.
\end{flushleft}
\epsfxsize=5.7in\epsfbox{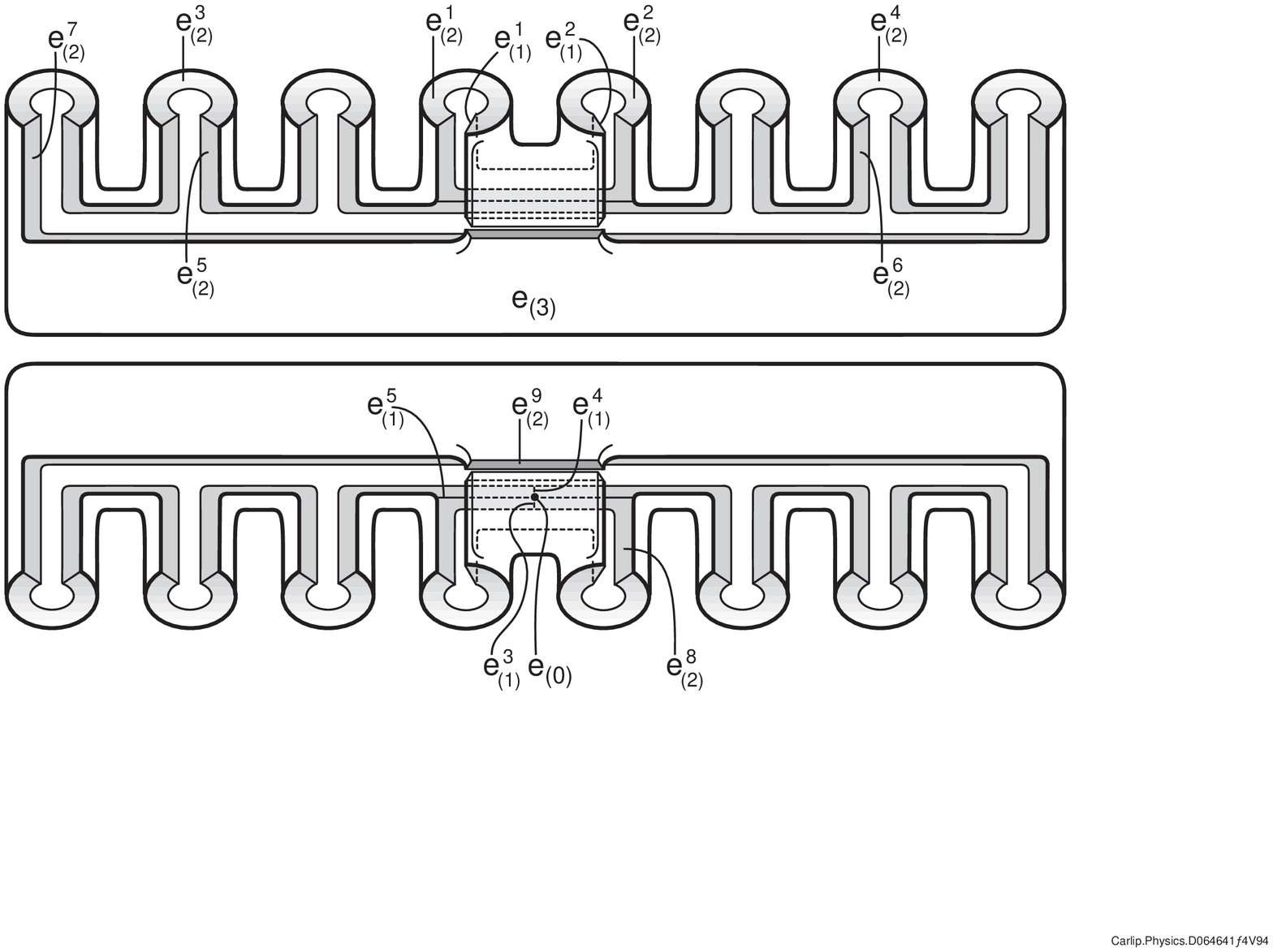}
\newpage
\begin{flushleft}
\small Figure 6.\ A two-parameter family of Reidemeister torsions $\tau(r,k)$
for different choices of the flat connection on $\partial M_2$; the range
shown is $1.1<r<12$, $.3<k<1.2$.
\epsfxsize=6in\epsfbox{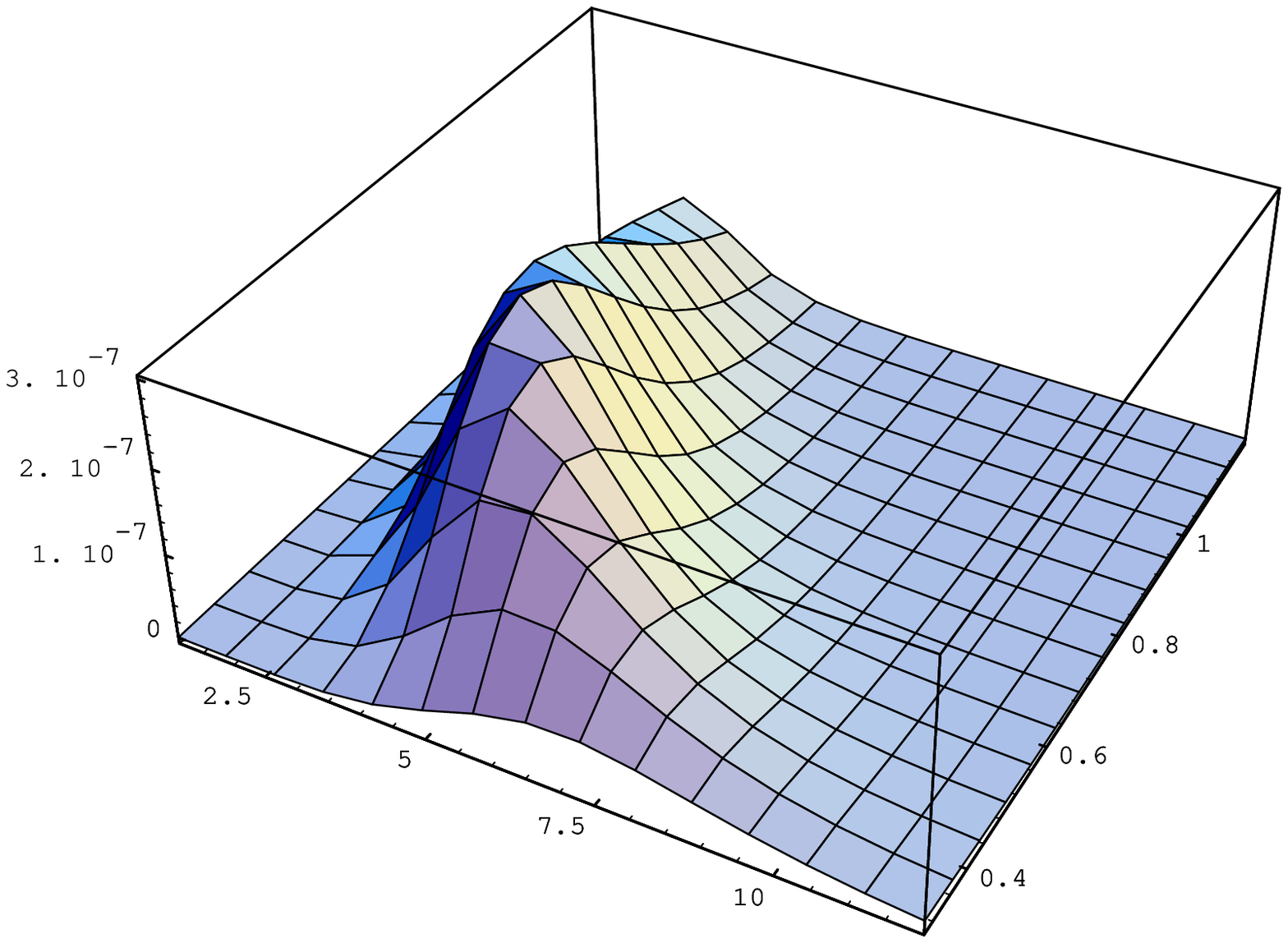}
\end{flushleft}

\end{document}